 \documentclass[11pt]{article}

\usepackage{amssymb}

\newtheorem{theorem}{$~~~~$ Theorem}[section]

\newtheorem{corollary}[theorem]{$~~~~$ Corollary}
\newtheorem{lemma}[theorem]{$~~~~$ Lemma}

\newtheorem{definition}[theorem]{$~~~~$Definition}

\def\nyp{\newpage }

\def\nor1{Normed$\{~2^{ \zzz \theta  \, )} ~$,$~\sqrt{~2^{ \zzz \theta  \, )}}~\}$}
\def\gggx{Available on-line  via Google.}

\def\xor2{Normed$\{ ~\sqrt{~2^{ \zzz \theta  \, )}}~,~2~ \} $}

\def\pag2{Page 2}

\def\zzz{~ \sharp ( ~ }

\def\f55{ \normalsize  \baselineskip = 1.8 \normalbaselineskip }

\def\f55{  \baselineskip = 1.1 \normalbaselineskip } 
\def\g55{  \baselineskip = 1.0 \normalbaselineskip } 
\def\s55{ \baselineskip = 1.0 \normalbaselineskip } 

\def\f55{  \baselineskip = 0.7 \normalbaselineskip }

\begin{document}

\title{Implications of the Trivers-Willard Sex Ratio
Hypothesis for Avian Species and Poultry Production,
And a Summary of the Historic Context of this Research}

\def\beq{\begin{equation}}
\def\enq{\end{equation}}

\def\bel{\begin{lemma}}
\def\enl{\end{lemma}}

\def\bec{\begin{corollary}}
\def\enc{\end{corollary}}

\def\bed{\begin{description}}
\def\ennd{\end{description}}
\def\bee{\begin{enumerate}}
\def\ene{\end{enumerate}}

\def\bxbxd{\begin{definition}}
\def\bxbxdd{\begin{definition}}
\def\eedd{\end{definition}}
\def\bxbxdr{\begin{definition} \rm}
\def\bel{\begin{lemma}}
\def\enl{\end{lemma}}
\def\ent{\end{theorem}}

\author{  Dan E.Willard}


\date{State University of New York at Albany}

\maketitle

\setcounter{page}{0}
\thispagestyle{empty}

\normalsize

\baselineskip = 1.3\normalbaselineskip

\normalsize

\baselineskip = 1.0 \normalbaselineskip 
\def\bbint{\large \baselineskip = 1.6 \normalbaselineskip } 
\def\bbint{\large \baselineskip = 1.6 \normalbaselineskip }
\def\bbint{\normalsize \baselineskip = 1.3 \normalbaselineskip }


\def\bbint{\normalsize \baselineskip = 1.27 \normalbaselineskip }

\def\bbint{\large \baselineskip = 2.0 \normalbaselineskip }

\def\bbint{\normalsize \baselineskip = 1.25 \normalbaselineskip }
\def\bbina{\normalsize \baselineskip = 1.24 \normalbaselineskip }

\def\bbint{\large \baselineskip = 2.0 \normalbaselineskip }

\def\bbing{ }
\def\bbins{ }
\def\bbinm{ }

\def\bbint{\normalsize \baselineskip = 1.95 \normalbaselineskip }

\def\bbing{ }
\def\bbins{ }
\def\bbinm{ }

\def\bbint{\large \baselineskip = 2.3 \normalbaselineskip } 
\def\bbing{ }
\def\bbins{ }
\def\bbinm{ }

\def\bbint{\normalsize \baselineskip = 1.7 \normalbaselineskip } 

\def\bbint{\large \baselineskip = 2.3 \normalbaselineskip } 
\def\bbinm{ \baselineskip = 1.18 \normalbaselineskip }

\def\bbint{\large \baselineskip = 2.0 \normalbaselineskip } 
\def\bbing{ }
\def\bbins{ }
\def\bbinm{ }
\def\bbinr{ }

\def\bbint{\normalsize \baselineskip = 1.25 \normalbaselineskip }
\def\bbina{\normalsize \baselineskip = 1.24 \normalbaselineskip }
\def\bbinr{ \baselineskip = 1.3 \normalbaselineskip }
\def\bbing{ \baselineskip = 1.28 \normalbaselineskip }
\def\bbins{ \baselineskip = 1.21 \normalbaselineskip }
\def\bbinm{  }

\def\ftl{ \baselineskip = 1.5 \normalbaselineskip }

\bbint

\parskip 5 pt

\begin{abstract}
\large
\baselineskip = 1.45 \normalbaselineskip

 At a theoretical level, the Trivers-Willard Sex Ratio
Hypothesis applies to both avian species and mammals.
This article, however,  conjectures that at the statistical level,
sex ratio effects are
likely to produce 
sharper numerical variations among
birds than among mammals.
We explain 
this greater statistical variation
should likely
have beneficial
implications for increasing the efficiency 
of world-wide
poultry egg (and perhaps  also meat)
production. 

\end{abstract}



\bigskip
\bigskip
\bigskip

\large
{\bf Keywords and Phrases:}
\normalsize
Fisher's Sex Ratio Theory, the Trivers-Willard Hypothesis,
Sex Determination Controls among Avian Species,
Enhancing
 Poultry Egg Production.

\bigskip
\bigskip
\bigskip

{\bf Mathematics Subject Classification:}
92D15; 92D25; 92D40; 92D50 

\bigskip
\bigskip
\bigskip

{\large  \bf Comment:} 
A 30-minute 
invited
talk summarizing the contents of this
article was  presented on June 18, 2017 at the
NEEPS-2017 Conference at Binghamton University. 
The Version-1 June 27 (2017)
draft of this
 paper was similar to
the 30-minute talk I gave at  Binghamton University,
and it was disseminated nine days after this
talk.  This second draft
is more detailed 
and polished
because it
contains
Section 4's additional evidence.


\bigskip

 \bigskip
%
%
%
%
%
%


\def\ww22{\normalsize \baselineskip = 1.21\normalbaselineskip \parskip 4 pt}
\def\bb22{\normalsize \baselineskip = 1.19\normalbaselineskip \parskip 4 pt}
\def\zz22z{\normalsize \baselineskip = 1.19 \normalbaselineskip \parskip 3 pt}
\def\xx22{\normalsize \baselineskip = 1.17\normalbaselineskip \parskip 4 pt}
\def\vx22s{\normalsize \baselineskip = 1.16 \normalbaselineskip \parskip 3 pt} 
\def\vv22{\normalsize \baselineskip = 1.17 \normalbaselineskip \parskip 3 pt} 
\def\aa22{\normalsize \baselineskip = 1.15 \normalbaselineskip \parskip 3 pt} 
\def\g55{  \baselineskip = 1.0 \normalbaselineskip } 
\def\s55{ \baselineskip = 1.0 \normalbaselineskip } 
\def\sm55{ \baselineskip = 0.9 \normalbaselineskip }

\vspace*{- 1.0 em}

\def\waw11{\normalsize \baselineskip = 1.72\normalbaselineskip}
\def\waw11{\normalsize \baselineskip = 1.12\normalbaselineskip}
\def\waw11{\normalsize \baselineskip = 1.85\normalbaselineskip}

\def\waw11{\normalsize \baselineskip = 1.45\normalbaselineskip}

\def\waw11{\normalsize \baselineskip = 1.7\normalbaselineskip}

\def\waw11{\normalsize \baselineskip = 1.12\normalbaselineskip}

\def\g55{  \baselineskip = 1.50 \normalbaselineskip } 
\def\s55{ \baselineskip = 1.50 \normalbaselineskip } 
\def\sm55{ \baselineskip = 1.5 \normalbaselineskip }

\def\g55{  \baselineskip = 1.50 \normalbaselineskip } 
\def\s55{ \baselineskip = 1.50 \normalbaselineskip } 
\def\sm55{ \baselineskip = 0.9 \normalbaselineskip }

\def\aa22{\normalsize  \waw11 \parskip 6 pt} 
\def\bb22{\normalsize  \waw11 \parskip 5 pt}
\def\ww22{\normalsize \waw11 \parskip 4 pt}
\def\vv22{\normalsize  \waw11 \parskip 3 pt} 
\def\tt22{\normalsize  \waw11 \parskip 2 pt} 

\def\g55{  \baselineskip = 1.0 \normalbaselineskip } 
\def\b55{  \baselineskip = 1.0 \normalbaselineskip } 
\def\s55{ \baselineskip = 1.0 \normalbaselineskip } 
\def\sm55{ \baselineskip = 0.9 \normalbaselineskip }

\def\mal{ \bf  }
\def\nal{\mathcal}

\def\cvrew{ \baselineskip = 1.6 \normalbaselineskip \parskip 3pt }

\def\ttt2c{ }
\def\tttc{ }

\def\tttc{\tiny \baselineskip = 0.8 \normalbaselineskip  \parskip 0pt }
\def\ttt2c{\tiny \baselineskip = 0.7 \normalbaselineskip  \parskip 0pt }
\def\tttc{ \baselineskip = 2.1 \normalbaselineskip  \parskip 5pt }
\def\ttt2c{ \baselineskip = 2.1 \normalbaselineskip  \parskip 5pt }

\def\tttc{ \baselineskip = 1.15 \normalbaselineskip  \parskip 5pt }
\def\ttt2c{ \baselineskip = 1.15 \normalbaselineskip  \parskip 5pt }

\def\tttc{ \baselineskip = 1.12 \normalbaselineskip  \parskip 4pt }
\def\ttt2c{ \baselineskip = 1.12 \normalbaselineskip  \parskip 4pt }

\def\tttc{ \baselineskip = 1.14 \normalbaselineskip  \parskip 3pt }
\def\ttt2c{ \baselineskip = 1.14 \normalbaselineskip  \parskip 4pt }

\def\cvt{ \baselineskip = 0.98 \normalbaselineskip }
\def\cv9{ \baselineskip = 0.99 \normalbaselineskip }
\def\cvs{ \baselineskip = 1.0 \normalbaselineskip }
\def\cvl{ \baselineskip = 1.0 \normalbaselineskip }
\def\cvh{ \baselineskip = 1.03 \normalbaselineskip }
\def\cvg{ \baselineskip = 1.00 \normalbaselineskip }

\def\cvt{ \baselineskip = 1.6 \normalbaselineskip }
\def\cv9{ \baselineskip = 1.6 \normalbaselineskip }
\def\cvs{ \baselineskip = 1.6 \normalbaselineskip }
\def\cvl{ \baselineskip = 1.6 \normalbaselineskip }
\def\cvh{ \baselineskip = 1.6 \normalbaselineskip }
\def\cvg{ \baselineskip = 1.6 \normalbaselineskip }
\def\cvb{ \baselineskip = 1.6 \normalbaselineskip }
\def\cvnew{ \baselineskip = 1.6 \normalbaselineskip }
\def\cvmew{ \baselineskip = 1.6 \normalbaselineskip }
\def\cvwew{ \baselineskip = 1.6 \normalbaselineskip \parskip 5pt }
\def\cvrew{ \baselineskip = 1.6 \normalbaselineskip \parskip 3pt }

\def\cvt{ \baselineskip = 1.22 \normalbaselineskip }
\def\cv9{ \baselineskip = 1.22 \normalbaselineskip }
\def\cvs{ \baselineskip = 1.22 \normalbaselineskip }
\def\cvl{ \baselineskip = 1.22 \normalbaselineskip }
\def\cvh{ \baselineskip = 1.22 \normalbaselineskip }
\def\cvg{ \baselineskip = 1.22 \normalbaselineskip }
\def\cvb{ \baselineskip = 1.22 \normalbaselineskip }
\def\cvnew{ \baselineskip = 1.4 \normalbaselineskip }
\def\cvmew{ \baselineskip = 1.35 \normalbaselineskip }
\def\cvwew{ \baselineskip = 1.4 \normalbaselineskip \parskip 5pt }
\def\cvrew{ \baselineskip = 1.22 \normalbaselineskip \parskip 3pt }

\def\cvt{ }
\def\cv9{ }
\def\cvs{ }
\def\cvl{ }
\def\cvh{ }
\def\cvg{ }
\def\cvb{ }
\def\cvnew{ } 
\def\cvmew{ }
\def\cvwew{ }
\def\cvrew{ }

\def\fend{ 

\medskip -------------------------------------------------------}

\def\g55{  \baselineskip = 1.0 \normalbaselineskip } 
\def\s55{ \baselineskip = 1.0 \normalbaselineskip } 
\def\sm55{ \baselineskip = 1.0 \normalbaselineskip } 
\def\h55{  \baselineskip = 1.08 \normalbaselineskip } 
\def\b55{  \baselineskip = 1.1 \normalbaselineskip } 

\normalsize

\baselineskip = 1.85 \normalbaselineskip





\parskip 2pt

\vspace*{- 1.0 em}



\baselineskip = 1.04 \normalbaselineskip 
\parskip 2pt

\baselineskip = 0.96 \normalbaselineskip 

%
\baselineskip = 2.16 \normalbaselineskip 
\baselineskip = 2.3 \normalbaselineskip 

\baselineskip = 0.95 \normalbaselineskip 
\baselineskip = 0.88 \normalbaselineskip 
\parskip 0pt
 
\def\fsp{ \baselineskip = 1.4 \normalbaselineskip}
\def\gvs{ }

\def\gvs{ \baselineskip = 1.0 \normalbaselineskip  \parskip 2pt}
\def\gvs{ \baselineskip = 2.0 \normalbaselineskip  \parskip 5pt}
\def\gvs{ \baselineskip = 1.0 \normalbaselineskip  \parskip 0pt}

\def\gvs{ \baselineskip = 1.6 \normalbaselineskip  \parskip 5pt}
\def\gvs{ \Large \baselineskip = 1.6 \normalbaselineskip  \parskip 7pt}
\def\gvs{ \large \baselineskip = 1.6 \normalbaselineskip  \parskip 6pt}
\def\gvs{ \normalsize \baselineskip = 1.6 \normalbaselineskip  \parskip 6pt}

\def\gvs{ \normalsize \baselineskip = 2.1 \normalbaselineskip  \parskip 7pt}
\def\gvs{ \normalsize \baselineskip = 1.8 \normalbaselineskip  \parskip    7pt}

\noindent

\newpage

\def\gvs{ \normalsize \baselineskip = 1.4 \normalbaselineskip  \parskip    5pt}
\def\gvs{ \normalsize \baselineskip = 1.44 \normalbaselineskip  \parskip    5pt}
\def\gvs{ \large \baselineskip = 1.44 \normalbaselineskip  \parskip    5pt}
\def\gvs{ \normalsize \baselineskip = 1.44 \normalbaselineskip  \parskip    5pt}\def\gvs{ \normalsize \baselineskip = 1.74 \normalbaselineskip  \parskip    5pt}
\def\gvs{ \normalsize \baselineskip = 1.44 \normalbaselineskip  \parskip 5pt}

\def\gvs{   \baselineskip = 1.74 \normalbaselineskip  \parskip    5pt}

\def\gvs{ \normalsize \baselineskip = 1.44 \normalbaselineskip  \parskip 5pt}
\def\gvs{ \large \baselineskip = 2.0 \normalbaselineskip  \parskip 5pt}
\def\gvs{ \Large \baselineskip = 2.0 \normalbaselineskip  \parskip 5pt}
\def\gvs{ \normalsize \baselineskip = 2.44 \normalbaselineskip  \parskip 5pt}
\def\gvs{ \normalsize \baselineskip = 2.04 \normalbaselineskip  \parskip 5pt}
\def\gvs{ \normalsize \baselineskip = 2.64 \normalbaselineskip  \parskip 5pt}
\def\gvs{ \Large \baselineskip = 1.6 \normalbaselineskip  \parskip 5pt}

\gvs

\footnotesize

\def\gvs{ }

\normalsize \baselineskip = 0.98 \normalbaselineskip
\normalsize \baselineskip = 1.0 \normalbaselineskip
\normalsize \baselineskip = 1.01 \normalbaselineskip

\def\gvs{ \normalsize \baselineskip = 1.25 \normalbaselineskip  \parskip 4pt}

\def\gvs{ \Large \baselineskip = 1.6  \normalbaselineskip  \parskip 6pt}
\def\gvs{ \normalsize \baselineskip = 1.6  \normalbaselineskip  \parskip 6pt}
\def\gvs{ \large \baselineskip = 1.6  \normalbaselineskip  \parskip 6pt}

\def\gvs{ \normalsize \baselineskip = 1.227 \normalbaselineskip  \parskip 3pt}
\def\gvs{ \large \baselineskip = 1.8  \normalbaselineskip  \parskip 6pt}

\def\gvs{ \normalsize \baselineskip = 1.5 \normalbaselineskip  \parskip 3pt}

\def\gvs{ \large \baselineskip = 2.1  \normalbaselineskip  \parskip 6pt}

\def\gvs{ \normalsize \baselineskip = 2.1  \normalbaselineskip  \parskip 6pt}


 \def\gvs{ \normalsize \baselineskip = 1.227 \normalbaselineskip  \parskip 3pt}

 \def\gvs{ \large  \baselineskip = 1.6 \normalbaselineskip  \parskip 5pt}

\def\gvs{ \Large  \baselineskip = 1.8 \normalbaselineskip  \parskip 5pt}
\def\gvs{ \LARGE  \baselineskip = 1.8 \normalbaselineskip  \parskip 5pt}
\def\gvs{ \normalsize  \baselineskip = 2.0 \normalbaselineskip  \parskip 5pt}

\def\gvs{ \Large  \baselineskip = 2.0 \normalbaselineskip  \parskip 5pt}

\def\gvs{ \large  \baselineskip = 2.2 \normalbaselineskip  \parskip 5pt}

\def\gvs{ \normalsize \baselineskip = 2.4  \normalbaselineskip  \parskip 6pt}

\def\gvs{ \normalsize \baselineskip = 2.6  \normalbaselineskip  \parskip 6pt}
\def\gvs{ \normalsize \baselineskip = 2.2  \normalbaselineskip  \parskip 6pt}
\def\gvs{ \normalsize \baselineskip = 1.8  \normalbaselineskip  \parskip 5pt}

\def\sgvs{ \small \baselineskip = 1.33  \normalbaselineskip  \parskip 1pt}
\def\tttc{ }
\def\ttt2c{ }

\def\gv2{ \normalsize \baselineskip = 1.30  \normalbaselineskip  \parskip 3pt}
\def\gvs{ \normalsize \baselineskip = 1.33  \normalbaselineskip  \parskip 5pt}

\def\gvs{ \normalsize \baselineskip = 2.33  \normalbaselineskip  \parskip 7pt}

\def\gvs{ \large \baselineskip = 2.63  \normalbaselineskip  \parskip 7pt}

\def\gvs{ \normalsize \baselineskip = 2.8  \normalbaselineskip  \parskip 7pt}
\def\gvs{ \normalsize \baselineskip = 2.0  \normalbaselineskip  \parskip 7pt}

\def\gvs{ \normalsize \baselineskip = 1.5  \normalbaselineskip  \parskip 7pt}

\def\gvs{ \large \baselineskip = 1.5  \normalbaselineskip  \parskip 7pt}

\def\gvs{ \normalsize \baselineskip = 1.4  \normalbaselineskip  \parskip 7pt}

\def\gvs{ \large \baselineskip = 1.7  \normalbaselineskip  \parskip 7pt}

\def\gvs{ \normalsize \baselineskip = 1.8  \normalbaselineskip  \parskip 7pt}

\def\gvs{ \normalsize \baselineskip = 1.4  \normalbaselineskip  \parskip 7pt}



\section{Introduction}

\label{ss1}

\gvs

The Trivers-Willard Hypothesis is an extension
of Fisher's Sex Equilibrium paradigm
\cite{Fish,Wilson}.
 Fisher
had
 noted
that essentially all vertebrate species will
invest a roughly equal amount of energy
raising 
 male and female offspring
because such 
a paradigm
 corresponds to the favored 
stabilizing
equilibrium
state that Darwinian Evolution  
gravitates towards
 over an extended
period of time. The Trivers-Willard Hypothesis (TW) was based on
the observation that some mating couples 
provide a setting (or have genes)
more useful to male reproductive success, while others
will be more supportive
to a  female's
reproductive success.
TW predicted that it would be
useful, accordingly,
 for Darwinian Evolution to attempt to guess
which
settings
 are
more
useful to which sex's reproductive
success,  and then 
to
adjust the sex ratio to reflect the
strategy that maximizes long-term reproductive success. 
This prediction will be called the {\it Generalized Version}
of the Trivers-Willard Hypothesis.

I conceived of this version of the TW during a 25-minute 
bicycle ride, after taking
in the Spring of 1968,
 a philosophy course at SUNY Stony Brook.
Its reading material included the Desmond Morris book
entitled {\it The Naked Ape} \cite{Naked}. During that fateful
bicycle ride, I also reconstructed Fisher's general sex ratio
paradigm (without knowing that Fisher conceived  of this 
idea 
almost forty
years earlier).  I also noticed
that the preceding ``Generalized'' theory  would imply that a
mating couple should be
more likely
to produce male offspring
 when they are in better health condition  
(or have an above average nutritional
diet).  
This latter
observation
 was a consequence of the well known anthropological
observation that competitive success benefits the 
reproduction rate of a male more than a female because a
male is capable of 
inseminating  simultaneously several females.

At the conclusion of this 1968 bicycle ride, I was left with a
quandary as to what to do next?  This is because I was uncertain
what part of the ``Generalized'' theory was new,
if any part of it was actually new?
There were several occasions in the past when I developed
theories that I learned,
 later,
 others had 
discovered substantially
 earlier
\footnote{\fsp 
For example during a bicycle ride in 1964, I reconstructed
many of the famous integration formulas, taught in a
Freshman Calculus course, without knowing about the famous
mathematical research done 200 years earlier.  
The resulting series of brain-storming thoughts actually
resulted in my discovering of Euler's irrational number
of ``e'', without knowing what Euler had 
done.
Humorously,
I naively called this number  ``w'' (for
``Willard's constant) before learning,
to my naive 
teen-age
chagrin, that Euler
discovered
approximately 
two centuries earlier
a
comparable
 irrational number he called 
 ``e''.} .
In the end, I decided to make no effort to publish the
Generalized Theory in 1968 because I thought it likely
that either someone had thought of a similar idea earlier, 
or it would be difficult to persuade the academic community
that these principles
were
 sound.
(My reluctance to
publish this  ``Generalized'' theory was further
amplified
by two family tragedies, where my father experienced a heart attack
and my mother was diagnosed with cancer during that same year of
1968. My mother was diagnosed with cancer,
actually,
 only a few weeks after
my bicycle ride.  I doubt I would have had the temperament to
discover the General Theory, had the chronology of these two
events been 
physically
 reversed.)

The remainder of the history behind the TW discovery has been told
by Robert Trivers. In 1970, I attended Harvard University as a graduate
student and audited a course taught by Irven Devore, where Robert
Trivers was a teaching assistant. In one lecture during that course,
Trivers reviewed Fisher's Sex Ratio Principle, and he mentioned in
a subsequent second lecture that it was  known that
upper class income families 
were statistically more likely to have male offspring.
I was delighted 
when I heard
 the first of Trivers's two lectures
because I had inadvertently reconstructed Fisher's
50/50 sex ratio principle during my
earlier
 1968 bicycle ride.
I explained to Trivers, 
subsequently
after his second lecture, my explanation for the
statistical 
paradigms
he mentioned. Trivers warmly encouraged me to publish this result. 

Distracted by the illness of both my parents, as well as
the burden of preparing for Harvard's  notoriously hard
Ph D Qualifying exam
in Mathematics,
 I did not pursue the TW project
further. As a 
consequence, Bob Trivers
helpfully
wrote up the manuscript of 
what would ultimately become 
our
announced 
result.
This
joint
paper \cite{TW73}
 did
later become a classic article,
according to
the MacArthur award-winning philosopher
Rebecca
 Goldstein \cite{Go12},
within
sociobiology's broader 
and ever-expanding
literature. 

A period of 49 years has now transpired between the current date
and  the occasion when I took that fateful 25-minute bicycle ride
in 1968.  It is accurate to state that no single 25-minute
investment 
of my time
(conjoined with Robert Trivers
excellent and
 meticulously  diligent
 write-up of our
joint paper \cite{TW73} )
did
produce
 a greater  
impact
 on the academic community
from
 my
on-going
research.
Thus on 23 June 2017, Google Scholar recorded
that there were 3,338 citations to \cite{TW73}'s
research, {\it  except for one 
hilarious
 error,}
 that deserves
to be in a Woody Allen movie. 
It will make
the
academic community laugh at the 
silly
 incompetence of the
careless
computer
engineers, who 
embedded an
amusing, almost schizophrenic bug into 
what was supposed to be
their   
sagacious
Google-Scholar software
\footnote{\fsp
The 
amusing  error is that neither Dan Willard nor Robert Trivers
were listed by 
Google-Scholar as the authors of the article 
\cite{TW73}, as recently
as 23 June 2017.
Instead, ``James A. McKanna'' is listed as the author
of an article with the same title and page numbers as
\cite{TW73}. This error occurred because {\it Science Magazine}
listed,
in 1973,
 the authors of its articles on 
the specified article's
last page (rather than on its first page). Moreover,
McKanna's 1973
paper
 ended on the first column of the same page 90,
whose second and third columns were occupied by the TW article.
Thus, the 
supposedly sagacious
Google-Scholar software had gone amusingly schizophrenic,
when it tried to guess who was the 
actual
 author of this particular article  with 
an unusual quantity of 3,338
 citations ? (This error persisted during all the Winter and Spring months
of the year 2017.
In fairness to Google,
their error was 
corrected on June 26, 2017, shortly after
I gave my June 18 talk at NEEPS-2017. 
We are not sure exactly why, but
Google Scholar's 
software has made
this 
persistent
error and then corrected it
repeatedly,
on several occasions, during the last few
years.) } .

In any case. this
little
 {\it ``Google Bug''}  is of small importance because 
the significance of the TW article
is well known.
For 
instance,
 there have been
four 
recent
mentions of this article in the
popular news media \cite{econ,nyt1,nyt2,psych}.
The latter 
has
included one year-2017 article in the 
{\it Sunday Week in Review} section of the
{\it New York Times} \cite{nyt2} .


Our purpose in the current short note will be to achieve
three goals. The first will be to suggest that avian species are likely
to follow the predictions of the TW hypothesis with greater
statistical accuracy than do mammals. A second objective
will be to suggest that this prediction is likely to
have
beneficial
implications in enhancing the efficiency
of poultry farms. A third goal, confined to  \textsection \ref{sect55},
will be to provide the reader with a brief summary of my
research into symbolic logic and 
into
G\"{o}del's Incompleteness theorem. This research has, traditionally,
been treated as a subfield of mathematics and philosophy.
But as we shall explain,  it also has
nontrivial
implications
for anthropology and psychology,
as well.


%
%
%


\section{A New Amendment to the Trivers-Willard Hypothesis for Avian
Species}
 
Our interest in
applying
 the TW hypothesis to avian
species was 
initially
stimulated by an article by Nancy Burley 
\cite{bur81yyyb}.   
It studied the
behavior of
 Australian Zebra Finches, and found that
their propensity to produce male offspring would be enhanced
if
colored
 bands were attached to their legs that made the males look
 more attractive and the females less attractive.
The
 reverse sex ratio would
be produced
if the bands had polar opposite
types of sexual attraction 
features.


I had not predicted such an effect.
However with retrospection,
I do have an interesting
explanation as to why avian species 
seem to
 follow
the predictions of the  Generalized   TW Hypothesis with more
statistical accuracy than mammals.


It is because
  sperm type determines the
sex of mammal offspring, while it is unfertized egg type that
is the control agent among birds.  
In particular, it is
 well
 known
that it is 
%
%
{\it the type},
X or Y,
of sperm
which
 determines
%
%
the sex among mammals
(e.g. an XX offspring is a daughter and an XY offspring is a son).
In contrast among birds, a ``ZZ'' genetic mix produces a male offspring,
and a ``ZW'' mixture corresponds to a female.  The latter implies
that it is the {\it unfertilized egg} (rather than donated sperm) that
functionally
determines the sex type for a bird.

Among both mammals and birds, the TW Hypothesis predicts that 
Darwinian Evolution has an incentive to guess which sex of
offspring is likely to produce more grandchildren for a mating
couple.  The engine, however, to determine which type of
sperm will first reach the egg is 
complicated, when a male mammal
donates several million 
competing
sperms,
at once.
 The comparable engine for
sex ratio
determination
 among
birds is, presumably, 
much
simpler 
because only a small number of
``Z'' or ``W'' unfertized eggs are deposited by the
mother for the purposes of being fertilized by a male.  

This distinction suggests  it will 
likely be {\it
substantially 
easier} for avian species to gain
full
 dexterous control over the
sex of their offspring than the analogous 
paradigm,  occurring
among mammal species.

Our 
suggested
amendment to the TW Hypothesis 
will
probably
 be very difficult to empirically
check for
its correctness. It would require a meticulous study
that compares various species of mammals to sundry species of
birds. It would, however, be theoretically interesting
if Avian species were found to obey the 
predictions of the
TW Hypothesis with greater 
statistical levels of
accuracy than 
among
mammals.
Moreover, the next section will suggest that our predictions,
if correct, could increase the 
world-wide
efficiency of poultry
egg production.

\section{ Poultry Farms}

It is well known that the efficiency of Poultry Farms
shall
increase
if more female chickens 
are born.  In that case,
egg production
will quickly increase, and also
poultry
 meat production
should
 also 
likely
  increase, somewhat.


It is apparent that if chickens 
do
function
 similarly to 
Australian Zebra Finches, then a mating couple will produce
more female offspring, if they are artificially endowed with female
color features. 

Moreover, a large variety of other techniques are likely 
to be also
available for
influencing the sex of offspring.  For instance, if an excess of male
rooster-like
sounds were pumped into a chicken farm then it is probable that more
females will be born (because the illusion of an excess
supply
 of males
would
have been
temporarily created).

One drawback of such strategies is that the inbred supply of farm
animals will naturally evolve in a direction 
{\it that is 
 exactly the
opposite} to a farmer's 
intentions.
This is because the classic 
local farm animal population will 
degrade,
spontaneously, 
in a direction
towards a 50/50
sex
 ratio, according to Fisher's
Sex Equilibrium  argument \cite{Fish}.

A useful remedy  is for 
a concerned
 farmer to keep a log
of which chickens come from a genetic 
lineage
producing more female
offspring --- and 
to encourage
those particular chickens to breed.

Unfortunately, such a log would 
require
a labor-intensive effort to
maintain, thus undermining its cost effectiveness. Fortunately,
there is a solution to this challenge in the modern computer age.
A unique 
computerized
bar-code name identifier could be attached to each chicken,
and a robot could ascertain that the correct 
genetic line of
chickens are breeding.

In other words, we are suggesting that a computerized algorithm could
maintain some type of desired protocol to enhance the ratio of female
offspring, and that this protocol will probably be cost effective in
the new age of computerized robotics that is now emerging. 
In any case, it is evident that the efficiency of poultry production
should be increased if farmers could gain better control of the
sex of raised chickens.

\section{Supporting Data}

\label{ss4}

A diversity of articles, by several authors,  will probably
be necessary to confirm our twin conjectures, suggesting that:
\bed
\item{ A. }
 Avian species do follow the predictions of the 
Trivers-Willard hypothesis with greater accuracy than do mammals.
\item{ B. } This paradigm can increase the
efficiency of poultry production.
\ennd
There is, however, adequate evidence in the
already-published literature to make these two conjectures
quite credible.

It firstly should be noted that there are some species
of laboratory animals, drawn from
the Vertebrate kingdom, where human 
experimenters
have gained 
essentially
100\%
control of the 
manipulated
sex of 
the studied
offspring. For instance, it has been
observed that reptiles and amphibians
 neither follow the mammal XX/XY or
the bird ZZ/ZW chromosomal method to control the sex of
offspring. Instead, both sexes have identical chromosome
structures, and 
it has been observed that laboratory scientists
can
  gain 100\%
control of the sex of offspring in 14 different genres of turtles
by changing the incubation temperatures for turtle
eggs.
(Thus,  \cite{BV79,BVB82,EJN94,VB82} observed
that a 25 Celsius incubation temperature produces an all-male
rate of offspring, while a 31 Celsius incubation temperature 
leads to 
an all-female population among 14 different
tested genres of turtles.)
A similar temperature-control effect has
also been observed
to occur in several
species of reptiles 
\cite{Bu80yyyr}.



Exact analogs of the preceding paradigm 
will not apply
to 
mating
chickens, since poultry uses a ZZ/ZW model, where a
ZZ animal is genetically a  male  and ZW is female.
It has, however,
 been observed that lowering the incubation
temperature does increase the frequency of female
births, partly because
 of sex-differential  mortality rates
and also because some genetically male (e.g. ZZ) chickens
possess a female anatomy and 
own
an 
observed
female-like
ability to lay eggs 
\cite{CSyyyt,GByyyt,Indyyyt,Myyyt},
if the incubating temperature
is lowered soon after the ZZ-egg is fertilized

There is a serious interest in the poultry industry to
increase the ratio of female births, as noted by
\cite{MyPetyyyf,Permyyyf,SRyyyf,Seyyyf,YDyyyf} 
among other
sources.  Several published articles have
studied the implications of TW hypothesis for avian
species
\cite{AVyyyb,bur81yyyb,bur86yyyb,DDByyyb,EGSyyyb,EELyyyb,GRHGyyyb,kilyyyb,kolyyyb,kdtmyyyb,PEyyyb,Syyyb,SNyyyb,SKRyyyb,WSyyyb,WByyyb},
and they
%
%
have noted it has had a 
documented
measured effect.
We suspect there in not yet enough
available
unambiguous
 evidence to determine whether our conjecture (A)
is precisely correct (e.g. that the TW hypothesis has
significantly
 greater
implications
among birds, than among mammals).
This 
particular
 topic
should be investigated 
in
much
greater
detail in the future.

If our conjecture (B) is correct (that poultry farms can
have their productivity increased by 
even
a few percentage points
through a better understanding of the implications of the
TW hypothesis) then the avian version of the TW hypothesis deserves
as much study as the  3,338 articles that have already 
examined
 its
implications for mostly mammals.

There is one particular experiment
that I would recommend be
undertaken.
It is known that the sex ratios within human societies change
in the aftermath of
wars \cite{Jyyyw,kyyyw}. {\it Would the same be true among
poultry species?} That is, what would happen if mating chickens
repeatedly heard sounds from an electronic speaker of tape-recordings
from a cock-fight?

The preceeding experiment is
a little
 awkward and embarrassing
to undertake,
 but the scientific information
that it 
supplied
could 
actually be quite
valuable
and sobering.

\section{Historic Context of this Research}

\label{sect55}

The first chapter of this article 
had
 mentioned that Willard's main contribution
to
the TW
 paper \cite{TW73} 
consisted of a flash insight that I developed
during a short bicycle ride
when I was 20 years old.
  The curiosity 
of many readers
may have been
stirred
 by this fact.
Some readers may, perhaps, 
begin to
 wonder what other intellectual
projects I have worked on,
subsequently,
 in the aftermath of
\cite{TW73}'s publication.

Essentially, my research has had two focal points. Prior to 1992, my
focus was on mainly classical topics concerning computer algorithm
design. My 
best known
work in this area consisted of a joint
study with Fredman to determine the optimal cost for
computerized sorting and related searching methodologies. Our joint
work showed that
the
 then-commonly-held presumption that
computerized
 sorting could
run  no faster than in $O(N~$Log$(N)~)$   
time
was incorrect (i.e. a 
theoretical speed-up for sorting and searching was demonstrated in
\cite{FW93,FW94} ). These two projects produced four papers (if one counts
separately their journal and conference 
publications).
It is reported in
Google Scholar that
 920  academic
citations to these four variations of our work
had subsequently appeared. 
Moreover, the
{\it 1991 Annual Report of the National Science Foundation}
 \cite{nsf}
cited
this particular  ``Fusion Tree" 
investigation
as the 
{\it chronologically first} among only six  projects
that were mentioned
in its 
1991
{\it Mathematics and Computer Science} section.
  
Starting in 1993, I started publishing papers 
\cite{ww93}-\cite{ww17}
about
G\"{o}del's 
historic
Incompleteness Theorem. 
G\"{o}del's work has traditionally been of interest to researchers
in fields as 
broadly
diverse as mathematics, philosophy and computing
(as a reader can 
quickly
surmise
by looking at any one of G\"{o}del's
biographies 
 \cite{Da97,Go5,Yo5} ). 
His   ``First Incompleteness Theorem''
indicated 
there existed no
systematic manner to categorize all the 
technically
true statements
{\it  in even
 the simplest
branch} of mathematics.
G\"{o}del's
``Second Incompleteness Theorem'' indicated conventional logical
systems are
also
 unable to confirm their own consistency, in a fully
formal sense 
\cite{End,Mend}.

There is no question that both these
 incompleteness
 results are rigorously correct,
but they raise the question about whether Darwinian Evolution might
favor the evolving of 
a more
advanced
specie of
 primate, 
 that finds it adaptive to
employ an
{\it unconventional mode of thought},
in order
 to
maintain
some type (?) of
specially
{\it modified 
knowledge} of its own consistency.
The latter 
topic
was the stimulus for our 
on-going
investigations in
\cite{ww93}-\cite{ww17}.
These articles proposed a variety of unconventional revisions
of arithmetic's formal axiomatic structure.
They found these delicate
revisions could preserve most of the pragmatic content of
traditional arithmetic, {\it while simultaneously providing }
at least 
some type of
philosophically
meaningful, albeit partially
diluted, formalized
appreciation of their
own internal
self consistency.

The best and indeed preferred paper to examine first,
 in the preceding
24-year long series of papers, is our final article
\cite{ww17}. The Remark 7.5 of \cite{ww17} mentions that we
suspect some variation of 
our
proposed ``IQFS'' formalism has
applications to anthropology, psychology and philosophy,
as well as to linguistics.
We do not suggest this paper is of easy reading.
A reader can, however, 
at least partially
appreciate 
\cite{ww17}'s
 gist, when its examination is conjoined
with also a reading of at least some select 
parts
of
the books \cite{Da97,End,Go5,Mend,Yo5}. 

We do not want to overstate this point,
but the Remark 7.5 
of \cite{ww17}
does indicate that our
newly proposed
``indeterminate
function
symbol''  $\theta$  should have implications
for each of the fields of
anthropology, psychology and philosophy,
as well as linguistics.
A wide spectrum of readers
is, thus, encouraged
to, at least, 
glance 
briefly
at 
\cite{ww17}'s discussion.

\smallskip

\section{Concluding Remarks }

\label{sect5}

The  main purpose of this article was to introduce our
proposed avian
 amendment
to the TW Sex Ratio Theory. This article  also included 
\textsection \ref{sect55}'s 
 brief
summary of our other research, during the last 44 years,
because 
we suspected  some readers would
find  its
short 
review to be 
informative, as well.

The main reason  the observations
in this short note
will
be
of  interest is because it is
possible
 that the world-wide egg (and plausibly also
poultry
 meat)
production 
could
undergo 
an approximate minimal
2-3 percentage 
or greater
 increase,
if the number of
born female chicklets is
significantly
 enhanced.
 Such a 
difference  will,
certainly,
 not resolve world-wide
famine challenges.
It  would, however,
 be a useful development,
beneficial to
mankind. 



\bigskip

\bigskip


{\bf ACKNOWLEDGMENT:} I thank Glenn Geher for his 
useful 
suggestion, conveyed 
on June 18 to me 
at the NEEPS-2017
conference \cite{glenn}, that my article
should 
also discuss
the relevance
of fascinating aspects of
 sex ratio effects,
that have been documented
for
 a variety of different
species of 
 turtles.



\begin{thebibliography}{99}

\small

 \baselineskip = 1.07  \normalbaselineskip 

\parskip 2 pt


\bibitem{AVyyyb}
C. Alonso-Alvarez and A. Velando,
``Female body conditon and brood sex ratio in
Yellow-legged Bulls Larus cachinnans'',
{\it British Ornithologists' Union}
145 (2003), pp.220-226.

\bibitem{Bu80yyyr}
J.J. Bull,
`` Sex determination in reptiles'',
{\it Quarterly Review in Biology,} 55 (1980), pp. 3-21.


\bibitem{BV79}
J.J. Bull and  R.V. Vogt,
``Temperature-dependent sex determination in turtles'',
{\it Science} (New Series Volume
 206), December 7, 1979, pp. 1186-1188.

\bibitem{BVB82}
J.J. Bull, R.V. Vogt and M. G. Bullmer,
``Heritability of  sex ratio 
in turtles with environmental sex
determination'',
{\it Evolution} 36:2  (1982), pp. 333-341. 


\bibitem{bur81yyyb}
N.  Burley,  
``Sex ratio manipulation and selection for 
attractiveness'', {\it Science}
(New Series, Volume 211),  February 13, 1981,  pp.721-722.

\bibitem{bur86yyyb}
N.  Burley,  
``Sex ratio manipulation in color-banded populations
of zebra-finches'',
{\it Evolution} 40 (1986), pp. 1191-1206.


\bibitem{CSyyyt}
J. Chue and C. A. Smith,
``Sex determination and sexual differetiation in the
avian model'',
{\it the FEBS Journal}
278 (2011), pp. 1027-1034. \gggx

\bibitem{Da97}
       {J. W. Dawson},
         {\it Logical Dilemmas: The life and work of
Kurt G\"{o}del},
   {AKPeters},
          1997

\bibitem{DDByyyb}
C. Dijkstra, S. Daan and J. B.  Burker,
``Seasonal variation in the sex ratios of kestrel broods'', 
{\it Functional Ecology} 4:2 (1990), pp.143-147.

\bibitem{econ}
{\it The Economist}, ``Sons and mothers'', July 11,  2009 (London), page
83 (no author given).

\bibitem{EGSyyyb}
H. Ellegren, L. Gustafsson and B. C. Sheldon,
``Sex ratio adjustment in relation to paternal attractiveness in a wild 
bird population'',
{\it Proceedings of National Academy of Sciences}
93 (1996), pp. 11723-11728.


\bibitem{EELyyyb}
S. T. Emlen. J. M. Emlen and S. A. Levin,
``Sex ratio selection in species with helpers-at-the-nest'',
{\it American Naturalist} 127:1 (1986) pp. 1-8.



\bibitem{End}
         {H. B Enderton},
 {\it   Mathematical Introduction  Logic},
        { Academic Press},
        2011

\bibitem{EJN94}
M. A. Ewert, D. R. Jackson and C. R. Nelson,
``Patterns of 
temperature-dependent sex determination in turtles'',
{\it Journal of Experimental Zoology} 270 (1994), pp. 3-15. 


\bibitem{Fish}
R. A. Fisher,
{\it The Genetical Theory of Natural Selection,}
Clarendon, Oxford,  1930.


\bibitem{FW93}
M. L. Fredman and D. E. Willard,
``Surpassing the information theoretic barrier with fusion trees'',
 {\it The
Journal of Computer and Systems Sciences} 
47 (1993), pp. 424-433.



\bibitem{FW94}
M. L. Fredman and D. E. Willard,
``Transdichotomous algorithms for minimum spanning trees
and shortest paths'',
 {\it The
Journal of Computer and Systems Sciences} 
48 (1994), pp. 533-551.


\bibitem{GRHGyyyb}
L. Gilbert, A. N. Rutstein, N. Hazon and J. A. Graves,
``Sex biased investments in yolk androgens 
depends on female quality and laying order in zebra finches'',
{\it Naturwissenschaften} 92:4 (2005), pp. 174-181.

\bibitem{glenn}
Glenn Geher, Private Communications on June 18 at the NEEPS-2017
conference, summarizing the literature about sex determination
in species of turtles.

\nyp

\baselineskip = 1.10  \normalbaselineskip 

\bibitem{Go5}
       R. Goldstein,
   {\it Incompleteness: The Proof and Paradox of Kurt G\"{o}del},
        {Norton} Press,
         2005.



\bibitem{Go12}
       R. Goldstein, private communications in 2012 during 
a visit by Goldstein to the SUNY-Albany campus. During those communications,
Goldstein told me that it was both her impression (and those of her husband
Steven Pinker) that the TW article \cite{TW73} played a major role in
the historic development of sociobiology because it was 
one of 
the first 
examples of a
sociobiology
 article making
purely
 theoretical predictions, that were later
statistically verified with
much
 corroborating evidence. 


\bibitem{GByyyt}
A. Goth and D. T. Booth,
``Temperature dependent sex ratio in a bird'',
{\it Biology Letters} (Royal Society), 1:1 ((March 22 2005), pp. 31-33. 
 \gggx

\bibitem{Indyyyt}
{\it The Independent},
``A drop in temperature can change the sex of chickens'',
an
author is unspecified for this
on-line
 news article, but it does quote
remarks by Prof Mark Ferguson from 
the
University of Manchester
(10 September 1997).  \gggx 


\bibitem{Jyyyw}
W. H. James, 
``The variations of human sex ratio at birth during and after
wars and their potential implications'',
{\it Journal of Theoretical Biology}
257:1 (2009), pp. 116-123,.

\bibitem{kyyyk1}
S. Kananzawa,
``Big and tall parents have more sons:
Further generalizatons of the Trivers-Willard hypothesis'',
{\it Journal of Theoretical Biology}
235 (2005), pp. 583-590.


\bibitem{kyyyk2}
S. Kananzawa,
``Beautiful parents have more daughters:
A further implication of the 
Trivers-Willard hypothesis (gTWH)'',
{\it Journal of Theoretical Biology}
244 (2007), pp. 133-140.

\bibitem{kyyyw}
S. Kananzawa,
``Big and tall soldiers are more likely to
survive a battle: A possible explanation for
the `returning soldier effect' on the
secondary sex ratio'',
{\it Human Reproduction} 22:11 (2007), pp. 3002-3008.


\bibitem{kyyyk3}
S. Kananzawa and N. L. Segal,
``Same-sex twins are taller than opposite-sex twins
(but only if breast-fed):
Possible evidence for sex bias in human breast milk'',
{\it Journal of Experimental Child Psychology},
available 
on-line 9 Janaury 2017 and to be published in
the near future, (Also accessible on 
Satoshi Kananzawa's home page.)

\bibitem{kilyyyb}
R. Kilner,
``Primary and scondary sex ratio manipulation
by zebra finches'',
{\it Animal Behavior}
56 (1998), pp.155-164.

\bibitem{kolyyyb}
J. Komdeur,
``Faculative sex ratio bias in the offspring of 
seychelles warblers'',
{\it Proceedings Biological Sciences}
263:1370 (1996), pp. 661-666 (Royal Society)


\bibitem{kdtmyyyb}
J. Komdeur, J. Daan, S. Tinbergen and C. Mateman,
``Extreme adaptive modification in sex ratio of 
seychelles warbler's eggs'',
{\it Nature}
385 (1997),
pp. 522-524.

\bibitem{Myyyt}
M. W. McDonald,
``Effects of temperature storage and age of
fowl eggs on hatchability and sex ratio,
growth and viability of the chickens'',
{\it Australian Journal of Agricultural Science}
11:4 (1960), pp. 664-672.
\gggx

\bibitem{Mend}
            {E. Mendelson},
          {\it Introduction to Mathematical Logic},
            { CRC Press},
        2010.




\bibitem{MyPetyyyf}
{\it My Pet Chicken} website (no author specified),
``How can I incubate eggs that will hatch female chicks only?'' \gggx

\bibitem{Naked}
D. Morris, 
{\it The Naked Ape}, Penguin Publishing House, 1967. 


\bibitem{nsf}
{\it National Science Foundation 1991 Annual Report},
``Mathematics and computer science section''.
(The Fredman-Willard ``fusion tree'' 
methodology of \cite{FW93,FW94}
was the chronolgically first
among
 six research projects mentioned in this section.)



\bibitem{nyt1} {\it The New York Times}, 
``Survival of species
linked to lopsided sex ratios'' (by Walter Sullivan),
front page of science section, February 16, 1981.


\bibitem{nyt2} {\it The New York Times}, 
``Does breast milk have a sex bias''
(by Nancy L Segal and Satoshi Kanazawa)
in Sunday Review section, January 21, 2017.
(See
\cite{kyyyk3}
 for a fascinating scientific study
that documents in more detail the
evdence germane to this newspaper article.) 



\bibitem{Permyyyf}
{\it Permies.com} website (no author specified),
``Can we hatch more hens and fewer roosters?'' \gggx



\bibitem{psych}
{\it Psychology Today}, ``10 politically incorrect truths
about human nature'' (by Alan S.  Miller and Satoshi Kanazawa),
August 2007,  pp. 88-95. (The Trivers-Willard Hypothesis was
``Politically Incorrect Truth \# 6";
see the articles \cite{kyyyk1,kyyyk2} for some scientific data
germane to this paper's  summarized discussion.)







\bibitem{PEyyyb}
C. B. Patterson and J. M. Emlen,
``Variation in nestling sex ratios in the yellow-headed blackbird'',
{\it The American Naturalist}
115:5  (1980) pp. 743-747,






\bibitem{RLB6}
J. R. Roche, J. M. Lee and D. B. Berry,
``Pre-conception energy balance and secondary sex
ratio -- Partial support of the Trivers-Willard
hypothesis in dairy cows'',
{\it Journal of Dairy Science}
89:6 (2006) pp. 2119-2125.


\bibitem{SRyyyf}
Y. Scott-Reid,
{\it Munchies} website,
``A scientist has figured out how to determine
chickens' sex before they hatch'',
April 8, 2015.
 \gggx




\bibitem{Seyyyf}
W. Seltzer,
``The method of controlling the sex of avian embryo,
improving embryo hachability and improving 
viability of the hatched chick'',
{\it United States Patent:} 2,733,482,
June 19, 1953.

\bibitem{Syyyb}
B. C. Sheldon,
``Recent studies of avian sex ratios'',
{\it Heredity} 80(1998), pp. 397-402.

\bibitem{SNyyyb}
E. Svenson and J. Nilsson,
``Mate quality affects sex ratio in blue tits'',
{\it Proceedings on Biological Sciences}
263:1368 (1996), pp. 357-361 (Royal Society)

\bibitem{SKRyyyb}
E. Szasz, D. Kiss and B. Rosivall,
``Sex ratio adjustments in birds'',
{\it Ornis Hungarica,} 20:1 (2012), pp. 26-36.


\bibitem{TW73}
R. L. Trivers and D. E. Willard,
``Natural selection of parental ability to vary sex ratio of
offspring'', {\it Science}
(New Series, Volume 179),  January 5, 1973, pp. 90-92,

\bibitem{VB82}
R. C. Vogt and J. J. Bull,
``Temperature controlled  sex determination in turtles: ecological
and behavorial aspects'',
{\it Herpetologica} 38:1 (1982), pp. 156-164.



\bibitem{WSyyyb}
S. A. West and B. C.  Sheldon,
``Constraints in the evolution of sex ratio adjustment'',
{\it Science} 295 
(March 1 2002), pp. 1685-1688.


\bibitem{WByyyb}
K. L. Wiebe and G. R. Bortolotti,
``Faculative sex ratio manipulation in American kestrels'',
{\it Behavioral Ecology and Sociabiology}
30:6 (1992), pp.379-386.


\bibitem{ww93}
D. E. Willard,
``Self-verifying axiom systems'', {\it
Computational Logic and Proof Theory:$~$
The   Third Kurt G\"{o}del
Colloquium}
(1993),
Springer-Verlag LNCS\#713,  325-336.


  

\bibitem{ww1}
D. E. Willard, ``Self-verifying  systems, the incompleteness
theorem and the tangibility reflection
principle'', in
{\it Journal of Symbolic Logic}
$~66~ (2001)\,$, pp. 536-596.


\bibitem{ww2}
D. E. Willard, 
``How to extend the semantic tableaux and
cut-free versions of the second
incompleteness theorem 
almost 
to 
Robinson's arithmetic Q'',
{\it Journal of Symbolic Logic}
$~\,67~ (2002)$, pp. 465--496.



\bibitem{ww5}
D. E. Willard, 
``An exploration of the partial respects in which an axiom
system recognizing solely addition as a total function can
verify its own consistency'', 
{\it Journal of Symbolic Logic} 70  (2005), pp. 1171-1209. 


\bibitem{wwapal}
D. E. Willard,  
``A generalization of the second incompleteness 
theorem and some exceptions to it''.
{\it Annals of Pure and Applied Logic}
141 (2006),
pp. 472-496.



\bibitem{ww6}
D. E. Willard, 
``On the available partial respects in which
 an axiomatization
for real valued  arithmetic can  recognize its 
consistency'', 
{\it Journal of Symbolic Logic} 71 (2006)
pp. 1189-1199.


\bibitem{ww7}
D. E. Willard,  
``Passive induction and a solution to a Paris-Wilkie 
question'',
{\it Annals of Pure and Applied Logic}
146(2007), 
pp. 124-149.


\bibitem{ww14}
         {D. E. Willard},
``On the broader epistemological
significance of self-justifying axiom systems'',
{\it Proceedings of 21st Wollic Conference},
Springer Verlag LNCS 8652 (2014), pp. 
{221-236}.

\bibitem{ww17}
D. E. Willard,  
``On  how the introducing of a 
 new $~\theta~$ function symbol
into arithmetic's formalism is germane
to devising axiom systems that can 
appreciate fragments of their own
Hilbert consistency'',
http://arxiv.org/abs/1612.08071 a 
Cornell Archives report.


\bibitem{Wilson}
E. O.  Wilson,
{\it Sociobiology: The New Synthesis},
Harvard University Press, 1975.


\bibitem{YDyyyf} 
B. Yilmaz-Dirkmen and S. Dirkmen,
``A morphometric method of sexing white layer eggs'',
{\it Brazilian Journal of Poultry Science}, 15:3 (2013), pp. 203-210.
\gggx  

\bibitem{Yo5}
         {P. Yourgrau},
              {\it A World Without Time: The Forgotten Legacy of
G\"{o}del and Einstein},
        {Basic Books},
           2005.






\end{thebibliography}
\end{document}